\documentclass[twoside,leqno,twocolumn]{article}

\usepackage[letterpaper]{geometry}

\usepackage{siamproceedings}
\usepackage{amsmath,amssymb}
\usepackage[T1]{fontenc}
\usepackage{amsfonts}
\usepackage{graphicx}
\usepackage{epstopdf}
\usepackage{enumitem}

\usepackage{caption}

\usepackage{algorithm}
\usepackage{algpseudocode}

\algtext*{EndIf}
\algtext*{EndFor}
\algtext*{EndWhile}

\ifpdf
  \DeclareGraphicsExtensions{.eps,.pdf,.png,.jpg}
\else
  \DeclareGraphicsExtensions{.eps}
\fi

\usepackage{booktabs}
\usepackage{tabularx}
\usepackage{mathtools}
\usepackage{array,booktabs,multicol}

\usepackage{microtype}

\usepackage[normalem]{ulem}
\usepackage{soul}

\newsiamremark{remark}{Remark}
\newsiamremark{hypothesis}{Hypothesis}
\crefname{hypothesis}{Hypothesis}{Hypotheses}
\newsiamthm{claim}{Claim}

\usepackage{thm-restate}
\newsiamthm{restatableproposition}{Proposition}
\newsiamthm{restatabletheorem}{Theorem}
\newsiamthm{restatablelemma}{Lemma}
\newsiamthm{restatablecorollary}{Corollary}

\newcommand{\DSFMSinkhornSweeps}{6}

\newcommand{\DSFMSeedRounds}{64}
\newcommand{\DSFMSlackSchedule}{(0,0.01,0.05,0.1,0.25,0.5,1,2,4)}
\newcommand{\DSFMLayerDepth}{8}
\newcommand{\DSFMPostDepth}{128}
\newcommand{\DSFMPostRounds}{4}
\newcommand{\DSFMExtraDepth}{4}
\newcommand{\DSFMLabelsPerVertex}{4}
\newcommand{\DSFMTerminalsPerRoot}{4}

\newcommand{\DSFMPackingMaxRoots}{12}
\newcommand{\DSFMPackingMaxPaths}{64}
\newcommand{\DSFMPackingMaxResources}{128}
\newcommand{\DSFMBnBNodeCap}{250{,}000}
\newcommand{\DSFMCongestionIterations}{16}
\newcommand{\DSFMWidenGrowth}{2}

\usepackage{amsopn}
\DeclareMathOperator{\diag}{diag}

\newcommand{\Nil}{\textsc{nil}}

\newcommand{\Infinity}{+\infty}

\newcommand{\card}[1]{\lvert #1\rvert}
\newcommand{\set}[1]{\{#1\}}

\newcommand{\RR}{\mathbb{R}}

\providecommand{\mA}{\ensuremath{\mat{A}}}

\providecommand{\mX}{\ensuremath{\mat{X}}}
\providecommand{\mY}{\ensuremath{\mat{Y}}}

\providecommand{\cA}{\ensuremath{\mathcal{A}}}

\providecommand{\cC}{\ensuremath{\mathcal{C}}}

\providecommand{\cE}{\ensuremath{\mathcal{E}}}
\providecommand{\cF}{\ensuremath{\mathcal{F}}}
\providecommand{\cG}{\ensuremath{\mathcal{G}}}

\providecommand{\cK}{\ensuremath{\mathcal{K}}}
\providecommand{\cL}{\ensuremath{\mathcal{L}}}

\providecommand{\cO}{\ensuremath{\mathcal{O}}}
\providecommand{\cP}{\ensuremath{\mathcal{P}}}

\providecommand{\cR}{\ensuremath{\mathcal{R}}}

\providecommand{\cT}{\ensuremath{\mathcal{T}}}
\providecommand{\cU}{\ensuremath{\mathcal{U}}}

\providecommand{\cY}{\ensuremath{\mathcal{Y}}}

\def\1{\bm{1}}

\def\mA{{\bm{A}}}

\def\mX{{\bm{X}}}
\def\mY{{\bm{Y}}}

\DeclareMathAlphabet{\mathsfit}{\encodingdefault}{\sfdefault}{m}{sl}
\SetMathAlphabet{\mathsfit}{bold}{\encodingdefault}{\sfdefault}{bx}{n}

\def\gG{{\mathcal{G}}}

\DeclareMathOperator*{\argmax}{arg\,max}

\newcommand{\GM}{\operatorname{GM}}

\newcommand{\pospart}[1]{[#1]_+}
\DeclareMathOperator{\mate}{mate}

\usepackage{bm}

\begin{document}

\newcommand\relatedversion{}

\title{\Large Dual-Slack Frontier Matching:
Parallel Approximate Maximum-Product Matching\relatedversion}

\author{
Marc A. Tunnell\thanks{
Purdue University, West Lafayette, IN, USA.
Email: \texttt{mtunnell@purdue.edu}.
}
\and
Oguz Selvitopi\thanks{
Lawrence Berkeley National Laboratory, Berkeley, CA, USA.
Email: \texttt{roselvitopi@lbl.gov}.
}
\and
Xiaoye Sherry Li\thanks{
Lawrence Berkeley National Laboratory, Berkeley, CA, USA.
Email: \texttt{xsli@lbl.gov}.
}
}

\date{}

\maketitle







\begin{abstract}
Maximum-product bipartite matching supports scaling and static pivot selection in sparse matrix preprocessing.
We present Dual-Slack Frontier Matching (DSFM), a deterministic, shared-memory parallel algorithm for approximate maximum-product matching.
Motivated by heavy matchings on small candidate supports, DSFM searches progressively exposed subgraphs ordered by fixed dual slack.
Slack-ordered certified completion guarantees a perfect matching whenever one exists, or a maximum-cardinality matching with a Hall certificate
otherwise.
Every returned perfect matching carries an a posteriori geometric-mean
quality bound; accumulated Hall obstructions can tighten this bound and improve the returned scaling.
We compare DSFM with Heavy-Weight Perfect Matching (HWPM) and exact MC64 assignment on 598 nonsingular matrices.
DSFM attains higher matching quality than HWPM over most quality thresholds.
On a set of large instances consisting of 37 matrices, DSFM achieves a $3.53\times$ geometric-mean speedup over MC64, compared to $2.29\times$ for HWPM.
\end{abstract}

\section{Introduction}
Weighted bipartite matching is a standard tool for permuting large entries onto the diagonal of a sparse matrix~\cite{Duff1999,Duff2001}.
Maximizing their product is equivalent to a weighted assignment problem~\cite{Kuhn1955}.
Such permutations and row--column scalings are good alternatives to partial pivoting to achieve highly scalable sparse unsymmetric LU decomposition.
In this paper, we use standard matching and graph-search terminology~\cite{CLRS2022}.

MC64's maximum-product option computes an optimal assignment and associated scaling~\cite{Duff2001}.
Its sequential execution can, however, become a preprocessing bottleneck in parallel sparse solvers~\cite{Azad2020}.
Heavy-Weight Perfect Matching (HWPM) addresses this limitation with a distributed-memory algorithm that first finds a perfect matching and then improves its weight through parallel searches for weight-increasing cycles of length four~\cite{Azad2020}.
Related approaches use shared-memory multi-source augmenting-path searches for cardinality matching~\cite{Azad2017} and diagonal-scaling preprocessing to reduce assignment instances, with a posteriori quality assessment~\cite{Sharify2013}.

Our approach is motivated by the empirical premise that a heavy-weight perfect matching often lies within a small, numerically significant selected subset of the matrix graph.
We make two principal contributions.
First, we introduce Dual-Slack Frontier Matching (DSFM), a shared-memory parallel algorithm that combines bounded, quality-guided search on progressively exposed subgraphs with structurally complete fallback.
DSFM returns a perfect matching with an a posteriori quality bound, rather than requiring exact maximum-product optimality, or a maximum-cardinality matching with a Hall certificate when no perfect matching exists.
Second, we develop accumulated Hall-obstruction corrections that tighten the quality bound and improve the returned scaling without changing the search slacks.
We prove that these corrections preserve dual feasibility and are never weaker than the strongest individual obstruction bound.
We establish correctness and work bounds and evaluate the resulting quality--performance tradeoff against MC64 and HWPM.

\section{Background}
Weighted bipartite matching is widely used to permute large entries onto the diagonal of sparse matrices and to support scaling, pivot selection, and preconditioning in sparse solvers~\cite{Duff2001,Duff2005}.
Let
$\mA\in\RR^{n\times n}$ be a sparse matrix with $a_{ij}$ denoting a nonzero entry.
We form the bipartite graph
\begin{equation}
    \gG=(\cC,\cR,\cE,\widetilde w), \qquad \card{\cC}=\card{\cR}=n,
\end{equation}
with column vertex $j \in \cC$, row vertex $i \in \cR$, and edge $(i,j) \in \cE$ for each nonzero of $\mA$ whose log magnitude is finite.
We denote by $\widetilde w$ the function for unnormalized edge weights and weight of $(i,j)$ is set as
\begin{equation}
  \widetilde w(i,j)=\log\lvert a_{ij}\rvert.
  \label{eq:log-weight}
\end{equation}
In the next subsection, we will normalize the weights and denote by $w$ the normalized form throughout the remainder of the paper.

When $\gG$ admits a perfect matching, maximizing the product of the selected magnitudes is equivalent to the maximum-weight assignment problem~\cite{Duff2001,Kuhn1955}:
\begin{equation}
  M^\star\in\argmax_{M\ \mathrm{perfect}} \widetilde w(M),
  \quad
  \widetilde w(M)=\sum_{(i,j)\in M} \widetilde w(i,j).
  \label{eq:max-assignment}
\end{equation}
If no perfect matching exists, the structural problem is instead to find a maximum-cardinality matching.
In this case, DSFM returns a Hall certificate, defined in subsection~\ref{sec:dual-cert}.

\subsection{Scaling and matching equivalence}
Sinkhorn--Knopp scaling alternates diagonal row and column normalizations of a nonnegative matrix, seeking a doubly stochastic scaling when one exists on its support~\cite{Knight2008,Sinkhorn1967}.
Let $r$ and $c$ be positive row and column scaling factors and, taking the absolute value function entrywise, define
\begin{equation}
    \mX = \diag(r) \lvert\mA\rvert \diag(c), \qquad \xi_{ij}=r_i \lvert a_{ij} \rvert c_j.
    \label{eq:X}
\end{equation}

\begin{algorithm}[!htb]
\caption{Sinkhorn--Knopp scaling}
\label{alg:sinkhorn}
\begin{algorithmic}[1]
\Procedure{Sinkhorn}{$\lvert\mA\rvert,k$}
  \State $r\gets\mathbf{1}$; $c\gets\mathbf{1}$
  \For{$k=1,\ldots,k$}
    \ForAll{$i\in\cR$ \textbf{in parallel}}
      \State $\displaystyle
        r_i\gets
        (\sum_{j:(i,j)\in\cE}\bigl\lvert a_{ij}\bigr\rvert c_j)^{-1}$
    \EndFor
    \ForAll{$j\in\cC$ \textbf{in parallel}}
      \State $\displaystyle
        c_j\gets
        (\sum_{i:(i,j)\in\cE}r_i \bigl\lvert a_{ij} \bigr\rvert)^{-1}$
    \EndFor
  \EndFor
  \State \Return $(r,c)$
\EndProcedure
\end{algorithmic}
\end{algorithm}

Each update normalizes the corresponding row or column sums of $\mX$ while holding the other scaling factors fixed.
Updates within each half-sweep are performed in parallel, with synchronization before the next half-sweep.
DSFM evaluates these normalizations in logarithmic coordinates for numerical stability; this changes only their numerical representation, not the scaling iteration defined above.

Under this diagonal scaling, the log-weight of edge $(i,j)$ becomes
\begin{equation}
  w(i,j)
  =
  \widetilde w(i,j)+\log r_i+\log c_j.\label{eq:w-tilde}
\end{equation}
Consequently, for every perfect matching $M$,
\begin{equation}
  w(M)
  =
  \widetilde w(M)+\sum_i\log r_i+\sum_j\log c_j.
\end{equation}
The final two sums are common to all perfect matchings, so diagonal scaling preserves their weighting relative to one another.
This conclusion holds for arbitrary positive scaling factors and does not require convergence of the iteration.

\subsection{Assignment duality and certificates}\label{sec:dual-cert}

The maximum-weight assignment in equation~\eqref{eq:max-assignment} admits the standard linear-programming dual~\cite{Duff2001,Kuhn1955}.
Associate a row potential $u_i$ with each $i\in\cR$ and a column potential $v_j$ with each $j\in\cC$.
The dual minimizes their total, subject to requiring that, for every edge, its row and column potential sum to at least its weight:
\begin{equation}
  \begin{aligned}
    \min_{u,v}\quad &\sum_i u_i+\sum_jv_j\\
    \text{s.t.}\quad
      &u_i+v_j\ge w(i,j),\qquad (i,j)\in\cE.
  \end{aligned}
  \label{eq:assignment-dual}
\end{equation}
The matching geometric means in the original and scaled matrices are,
\begin{equation}
  \begin{aligned}
  \GM_\mA(M)
  &=\bigl(\prod_{(i,j)\in M}\lvert a_{ij}\rvert\bigr)^{1/n}
  =\exp\!\bigl(\frac{\widetilde w(M)}{n}\bigr),\\
  \GM_\mX(M)
  &=\bigl(\prod_{(i,j)\in M} \xi_{ij} \bigr)^{1/n}
 =\exp\!\bigl(\frac{w(M)}{n}\bigr).
  \end{aligned}
  \label{eq:geomean}
\end{equation}

Any feasible pair $(u,v)$ provides an upper bound on the weight of a perfect matching.
For each edge $(i,j)$, define the reduced slack
\begin{equation}
  s(i,j)=u_i+v_j-w(i,j).
  \label{eq:slack}
\end{equation}
Dual feasibility is equivalent to nonnegative slack on every edge.
An edge is \emph{tight} when its slack is zero.
DSFM uses these slacks to sort candidate edges and to construct the a posteriori objective certificate described below.

In exact arithmetic, the dual surface constructed by DSFM is feasible.
For a perfect matching $M$, define
\begin{equation}
    s(M) = \sum_{(i,j)\in M}s(i,j).
\end{equation}
Let $\omega\ge0$ be a certified lower bound on the total fixed slack of
every perfect matching.
DSFM initializes $\omega=0$ and accumulates contributions from newly certified slack intervals during completion (Proposition~\ref{prop:accumulated-corrections}).
DSFM returns the objective certificate in the original log-weight coordinates,
\begin{equation}
    Q=(\widetilde w(M),\,s(M),\,\omega).
    \label{eq:objective-certificate}
\end{equation}
For a maximum-weight perfect matching $M^\star$, the common row and column
potential sums give
$w(M^\star)-w(M)=s(M)-s(M^\star)$.
Since $s(M^\star)\ge\omega$ and diagonal scaling preserves weight
differences,
\begin{equation}
    \widetilde w(M)
    \le \widetilde w(M^\star)
    \le \widetilde w(M)+s(M)-\omega.
\end{equation}
The corresponding geometric mean ratios satisfy
\begin{equation}
    \begin{aligned}
        \frac{\GM_\mA(M)}{\GM_\mA(M^\star)}
        &=
        \frac{\GM_\mX(M)}{\GM_\mX(M^\star)}\\
        &\ge
        \exp\!\biggl(-\frac{s(M)-\omega}{n}\biggr).
    \end{aligned}
\end{equation}
When $s(M)=\omega$, the certificate proves maximum-weight optimality,
even if the matched slack is positive.
A positive remaining gap bounds the possible objective loss but does not
establish suboptimality.
When $\omega=0$, the certificate reduces to the matched-slack bound.

If $\gG$ has no perfect matching, DSFM instead returns a structural certificate.
For $S\subseteq\cC$, let $N_{\gG}(S)$ denote its row--neighbor set.
Hall's theorem states that a perfect matching exists if and only if $\card{N_{\gG}(S)}\ge\card{S}$ for every $S\subseteq\cC$; a pair violating this condition, i.e.,
\begin{equation}
  H=(S,T),
  \quad
  T=N_{\gG}(S),
  \quad
  \card{T}<\card{S},\label{eq:hall-certificate}
\end{equation}
is a directly verifiable certificate of structural singularity~\cite{Hall1935}.
We refer to such a pair $H=(S,T)$ as a Hall certificate.

Throughout, whenever the numerical priorities specified for an operation do not uniquely determine a choice, DSFM resolves the remaining ties by a fixed total order on structural identifiers, including row, column, root, predecessor, and path identifiers as applicable.
This order is independent of parallel execution but depends on the row and column labeling of the input; permuting either may therefore change the selected matching.
Consequently, in exact arithmetic, DSFM produces the same result for a fixed labeled input across repeated executions and thread counts.
This convention applies to all nonunique choices below and is not repeated.

\section{Dual-Slack Frontier Matching Algorithm}
DSFM is a deterministic, shared-memory parallel algorithm that constructs a matching over progressively enlarged admissible subgraphs.
Rather than repeatedly searching the full matrix graph, it first concentrates combinatorial work on a small set of numerically promising edges and exposes additional alternatives as the search proceeds.

\begin{figure}[!htb]
    \centering
    \includegraphics{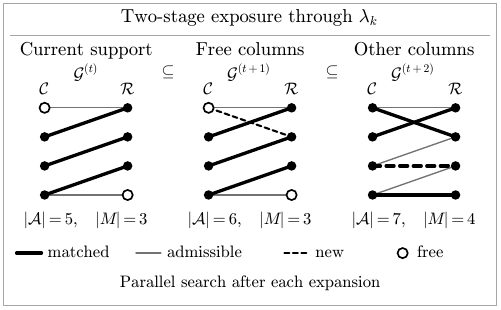}
    \caption{Two-stage exposure at one slack target $\lambda_k$.
    Thick edges show the matching after each search; dashed edges are newly exposed, and open vertices are free.}
    \label{fig:dsfm-staged-exposure}
\end{figure}

Sinkhorn sweeps followed by row--column potential updates produce dual-feasible potentials and nonnegative edge slacks.
These slacks remain fixed throughout the matching search, a policy we call \emph{fixed-slack search}.
A slack target $\lambda$ permits exposure of edges satisfying $s(i,j)\le\lambda$, equivalently those whose fixed-dual-scaled magnitudes $e^{-s(i,j)}$ are at least $e^{-\lambda}$.
The scheduled search uses increasing targets $0=\lambda_0<\cdots<\lambda_K$, with all tight edges included in the initial admissible graph.
Each expansion advances the admissible prefixes of selected columns; edges are never removed.
Writing $\cG^{(t)}$ for the admissible graph after exposure event $t$, we therefore obtain
\begin{equation}
    \cG^{(0)}\subseteq\cG^{(1)}\subseteq\cdots,
\end{equation}
where inclusion refers to edge sets and $\cG^{(0)}$ is the initial admissible graph.
Throughout, the vertex set is kept fixed.
Within each expanded column, newly exposed edges have larger fixed slack, and hence smaller dual-scaled magnitude, than previously exposed edges.
Figure~\ref{fig:dsfm-staged-exposure} illustrates two such expansions at a single slack target.

Parallel seeding and score-guided alternating-path searches augment the matching within the current admissible graph.
Independent path extensions are resolved by deterministic retention rules, and path-disjoint augmentations are committed together.
If a perfect matching is found, DSFM returns it with an objective certificate.
Otherwise, after the bounded search phases, certified completion continues the slack-ordered exposure and the exhaustive alternating search until it finds a perfect matching or a Hall certificate.
Fixed slack remains exactly aligned with the perfect-matching objective: for every perfect matching $M$,
\begin{equation}
  s(M) = \sum_i u_i+\sum_jv_j-w(M).
\end{equation}
Maximum-weight perfect matchings are therefore precisely those of minimum total fixed slack.
Low-slack exposure guides the search but does not establish global optimality: a later, higher-slack edge can enable a lower-total-slack reassignment.
The returned certificate bounds the possible remaining improvement over the full graph using the actual matched slack and accumulated obstruction correction.

\subsection{Algorithmic state and invariants}
We formulate the algorithm in the log-coordinates of the scaled matrix $\mX$ defined in~\cref{eq:X}.
The row and column potentials initialize to zero.
Each sweep first updates all row potentials with the column potentials held fixed, then updates all column potentials using the new row values:
\begin{equation}
  \begin{aligned}
    u_i&\gets
      \max_{j:(i,j)\in\cE}\{ w(i,j)-v_j\},\\
    v_j&\gets
      \max_{i:(i,j)\in\cE}\{ w(i,j)-u_i\}.
  \end{aligned}
  \label{eq:uv-updates}
\end{equation}
Updates within either half-sweep are independent and performed in parallel; the column half-sweep follows completion of the row half-sweep.
The implementation represents these scaled coordinates implicitly using the stored sparse weights and scaling factors; $\mX$ is never explicitly stored.

After these updates, the potentials remain fixed during the matching search.
In the scaled coordinates, the reduced slack is
\begin{equation}
   s(i,j)=u_i+v_j- w(i,j).
   \label{eq:scaled-slack}
\end{equation}
These fixed slacks determine the tight edges and the progressively enlarged low-slack admissible sets used by the search.
Objective-dependent comparisons continue to use the original log-weights $\widetilde w(i,j)$.

The row update makes the potentials dual feasible and ensures every nonempty row is incident on a tight edge.
It also gives $w(i,j)-u_i\le v_j^{\mathrm{old}}$ on every edge, so the subsequent column update satisfies $v_j^{\mathrm{new}}\le v_j^{\mathrm{old}}$.
For any edge attaining a row maximum, equality holds, so its column potential is unchanged.
The column update therefore preserves a tight incident edge for every nonempty row while ensuring the same property for every column.
After one complete sweep, the potentials are dual feasible in exact arithmetic and every vertex is incident on a tight edge.

Write $\cE_j=\set{(i,j)\in\cE:i\in\cR}$ and order its edges by nondecreasing fixed slack, resolving ties by the prescribed structural order.
For $\eta\ge0$, define
\begin{equation}
\cE_j^{\leq\eta}=\set{(i,j)\in\cE_j:s(i,j)\leq\eta}.
  \label{eq:column-slack-layer}
\end{equation}

Each admissible set $\cA_j\subseteq\cE_j$ is a prefix that includes all edges tied at an exposed slack value.
These prefixes persist and grow monotonically, defining
\begin{equation}
  \cA=\bigcup_{j\in\cC}\cA_j,
  \qquad \cG_{\cA}=(\cC,\cR,\cA,w).
  \label{eq:current-admissible-graph}
\end{equation}

An expansion at columns $S\subseteq\cC$ through slack $\eta$ replaces the old prefixes $\cA_j^-$ by
\begin{equation}
  \cA_j^+=
  \begin{cases}
      \cA_j^-\cup\cE_j^{\leq\eta}, & j\in S,\\
      \cA_j^-, & j\notin S.
  \end{cases}
  \label{eq:admissible-expansion}
\end{equation}
The new edges are $\Delta\cA_j=\cA_j^+\setminus\cA_j^-$ and $\Delta\cA=\bigcup_j\Delta\cA_j$; earlier edges remain admissible.
The search \emph{frontier}, in contrast, consists of active column vertices awaiting expansion.

\subsection{Supporting operations}

We use the following named operations in the algorithms presented in this section.
The descriptions and algorithm invocations below use the parameter values of the reference implementation.
Detailed pseudocode for these operations is omitted for brevity; their complete implementations are open-source and preserved as-used in this work at 
\href{https://doi.org/10.5281/zenodo.22761056}{doi.org/10.5281/zenodo.22761056}.

{\bf $\Call{InitializeDuals}{\gG}$} performs at most \DSFMSinkhornSweeps{} Sinkhorn sweeps (Algorithm~\ref{alg:sinkhorn}).
After each sweep, before proceeding to the next, DSFM selects in parallel one maximal entry from each column of the current scaled matrix $\mX$.
If these entries occupy distinct rows, DSFM invokes the optimality check described below using the current scaling.
The screening selection only triggers this check; its edges are not retained as a matching.

The optimality check applies one row--column potential sweep (equation~\eqref{eq:uv-updates}) to obtain dual-feasible potentials.
It then selects in parallel one tight incident edge for each column $j$, with row index
\begin{equation}
    i_j^\mathrm{tight}
    \in
    \set{i\in\cR:(i,j)\in\cE_j,\ s(i,j)=0}.
\end{equation}
Because these choices are made independently, different columns may select the same row.
If all selected row indices are distinct, the edges form the perfect matching
\begin{equation}
    M_0=\set{(i_j^\mathrm{tight},j):j\in\cC}.
    \label{eq:early-matching}
\end{equation}
Every edge of $M_0$ is tight under dual-feasible potentials, so $M_0$ is maximum-weight and initialization returns immediately with $(u,v,M_0)$.
Otherwise, the selection is discarded and the Sinkhorn iteration continues if sweeps remain.
If no early return occurs, initialization returns the final dual-feasible potentials with $M=\emptyset$ for the subsequent matching search.

{\bf $\Call{Seed}{\cG_{\cA},M}$} specializes frontier search to length-one augmenting paths from free columns to free rows.
Workers propose admissible edges in parallel across free columns, resolve conflicts deterministically by destination row, and commit a noncompeting subset in parallel without changing existing matched edges.
At most \DSFMSeedRounds{} rounds are performed; no matched-edge continuation, predecessor chains, or general path packing is needed.

{\bf $\Call{AugmentExpandedSet}{\cG_{\cA},M}$} first invokes $\Call{Seed}{\cG_{\cA},M}$ to capture inexpensive length-one augmentations.
If $M$ remains non-perfect, it invokes the general $\Call{FrontierSearch}{\cG_{\cA},M}$ procedure (Algorithm~\ref{alg:frontier}) with maximum depth \DSFMLayerDepth.

\subsection{Overall algorithm}\label{sec:overall}
DSFM seeks a perfect matching on a small, numerically selected support and stops once one is found, returning an a posteriori bound on the possible objective improvement over the full graph rather than requiring exact maximum-product optimality.
Its phase-synchronous execution parallelizes half-sweeps over rows or columns, prefix expansion and seeding across columns, search within each depth, and packing across conflict components.
Barriers separate dependent stages.
Selected vertex-disjoint paths are committed in parallel before another search begins.

Algorithm~\ref{alg:dsfm} uses the slack schedule
\begin{equation}
    \Lambda=\DSFMSlackSchedule.
\end{equation}
Fine early targets favor low-slack choices while many columns remain free; coarser later increments limit the number of exposure stages.
Exploratory tests showed little sensitivity to the precise targets with this progression; the sequence is empirical and is not claimed optimal.
A moderate terminal target limits the support handed to completion, which can expose further edges locally and therefore does not depend on this limit for structural completeness.

For each target $\lambda_k$, $k=1,\ldots,K$, let $\cU$ be the columns free at the iteration's start.
DSFM first expands their admissible prefixes through $\lambda_k$ and attempts augmentation when new edges appear.
If $M$ remains non-perfect, it expands $\cC\setminus\cU$ through the same target and repeats the attempt when new edges appear.
This supplies new root choices before matched-column alternatives and avoids the second expansion when the first search finds a perfect matching (Figure~\ref{fig:dsfm-staged-exposure}).
Advancing the target reflects exhausted search budgets, not proof that the current support lacks an augmenting path.

After exhausting $\Lambda$, DSFM performs up to \DSFMPostRounds{} additional searches with maximum depth $D=\DSFMPostDepth$.
Shallow searches ($D=\DSFMLayerDepth$ during exposure and after certified widening) limit synchronization and packing work but may leave more to completion; deeper searches permit longer alternatives but must offset their extra cost by reducing subsequent completion work.
A smaller deficit at handoff need not imply lower total runtime.
If $M$ remains non-perfect, $\Call{CertifiedCompletion}{\gG,\cA,M,\omega}$ (Algorithm~\ref{alg:certified}) finds a perfect matching or returns a Hall certificate.

After exhausting $\Lambda$, DSFM invokes the score-guided frontier search for up to \DSFMPostRounds{} additional rounds with maximum depth $D=\DSFMPostDepth$.
Shallow searches, using $D=\DSFMLayerDepth$ during scheduled exposure and after certified widening, favor short candidates and limit depth synchronization and packing work, but may leave more augmentations to completion.
A larger depth limit permits longer alternatives within the same support and may reduce the remaining matching deficit, at the cost of additional search and potentially more expensive packing.
A smaller deficit at handoff does not necessarily imply lower total runtime: the deeper search must offset its own cost.
If the matching remains non-perfect, we invoke $\Call{CertifiedCompletion}{\gG,\cA,M,\omega}$ (Algorithm~\ref{alg:certified}).
It either augments $M$ to a perfect matching or returns a Hall certificate (equation~\eqref{eq:hall-certificate}).

In Algorithm~\ref{alg:dsfm}, $\mate_M(v)$ denotes the unique vertex matched to $v\in\cC\cup\cR$ by $M$, or $\Nil$ if $v$ is free.
The separate output arrays $(u^{\mathrm{out}},v^{\mathrm{out}})$ are initialized from $(u,v)$ and updated in place by certified completion; they are omitted from procedure signatures and return tuples for brevity.

\begin{algorithm}[!htb]
\caption{Dual-Slack Frontier Matching}
\label{alg:dsfm}
\small
\begin{algorithmic}[1]
\Require $\gG$ and increasing slack schedule
         $\Lambda=(\lambda_0,\ldots,\lambda_K)$
\Ensure Matching $M$ with objective certificate $Q$ or Hall certificate $H$. \Comment{see equations~{\eqref{eq:objective-certificate}--\eqref{eq:hall-certificate}} for $Q$ and $H$.}
\Procedure{DualSlackFrontierMatch}{$\gG,\Lambda$}
  \State $(u,v,M)\gets\Call{InitializeDuals}{\gG}$
  \State $(u^{\mathrm{out}},v^{\mathrm{out}})\gets(u,v)$
  \State $Q\gets\Nil$; $H\gets\Nil$; $\omega \gets 0$
  \If{$M$ is perfect} \Comment{Early-exit triggered~\cref{eq:early-matching}}
      \State $Q\gets(\widetilde w(M),0,0)$
      \State \Return $(M,Q,H)$
  \EndIf
  
  \State $\cA_j\gets\cE_j^{\leq\lambda_1}$ for every $j\in\cC$
  \State \Call{AugmentExpandedSet}{$\cG_{\cA},M$}
   
    \For{$k\gets1$ \textbf{to} $K$}
      \If{$M$ is perfect} \textbf{break} \EndIf
      \State $\cU\gets\set{j\in\cC:\mate_M(j)=\Nil}$ \Comment{unmatched}
      \State Expand $\cA_j$, $j\in\cU$, through $\lambda_k$; record $\Delta\cA$
      \If{$\Delta\cA\ne\emptyset$}
        \State \Call{AugmentExpandedSet}{$\cG_{\cA},M$}
      \EndIf
      
      \If{$M$ is perfect} \textbf{break} \EndIf
      \State Expand $\cA_j$, $j\in\cC\setminus\cU$, through $\lambda_k$; record $\Delta\cA$
      \If{$\Delta\cA\ne\emptyset$}
        \State \Call{AugmentExpandedSet}{$\cG_{\cA},M$}
      \EndIf
    \EndFor
    \For{$\DSFMPostRounds$ \text{iterations}}
      \If{$M$ is perfect} \textbf{break} \EndIf
      \State $a\gets\Call{FrontierSearch}{\cG_{\cA},M}$
      \Comment{depth \DSFMPostDepth}
      \If{$a=0$} \textbf{break} \EndIf
    \EndFor
    \If{$M$ is not perfect}
      \State $(H,\omega)\gets\Call{CertifiedCompletion}{\gG,\cA,M,\omega}$
    \EndIf
  \If{$M$ is perfect}
      \State $Q\gets(\widetilde w(M),s(M),\omega)$ \Comment{see equation~\eqref{eq:objective-certificate}}
  \EndIf
  \State \Return $(M,Q,H)$
\EndProcedure
\end{algorithmic}
\end{algorithm}

\subsection{Returned scaling}\label{sec:returned-scaling}
The scaling associated with a maximum-product assignment need not be unique, and different choices can affect solver behavior~\cite{Hook2019}.
DSFM returns the combined scaling factors $\widehat r_i=r_i e^{-u_i^{\mathrm{out}}}$ and $\widehat c_j=c_j e^{-v_j^{\mathrm{out}}}$.
The separate output potentials copy $(u,v)$ at initialization and accumulate the obstruction corrections in Algorithm~\ref{alg:certified}; the search potentials remain fixed.
Proposition~\ref{prop:accumulated-corrections} shows that the output potentials are dual feasible and their total is smaller by $\omega$.
Thus, in exact arithmetic, $\mY=\diag(\widehat r)|\mA|\diag(\widehat c)$ has every entry at most one, and for a returned perfect matching $M$,
\begin{equation}
  \GM_{\mY}(M)
  =\exp\!\left(-\frac{s(M)-\omega}{n}\right).
  \label{eq:returned-scaling-quality}
\end{equation}
This is a factor $\exp(-(s(M^\star)-\omega)/n)$ of the largest matched geometric mean achievable for the same $M$ by diagonal scaling subject to the entrywise unit bound.
Optionally, after all corrections have been accumulated, one row--column sweep~\eqref{eq:uv-updates} over the full graph tightens the output potentials in $\cO(m+n)$ work.
The sweep preserves feasibility and cannot decrease any scaled magnitude relative to the corrected matrix, so~\eqref{eq:returned-scaling-quality} becomes a lower bound.
Neither operation changes the matching or the fixed slacks used during search and certification.

\section{Score-guided frontier search}\label{sec:frontier}
\begin{figure}[!htb]
    \centering
    \includegraphics{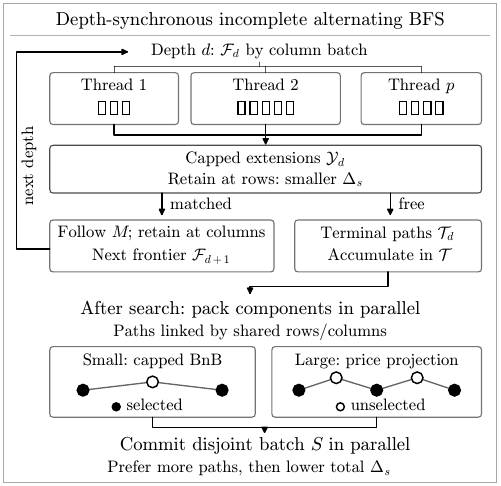}
    \caption{Parallel score-guided frontier search.
    Threads cooperatively perform a depth-synchronous, quality-filtered alternating BFS, generating and retaining bounded extensions at each depth.}
    \label{fig:dsfm-parallel-frontier}
\end{figure}

$\Call{FrontierSearch}{}$ performs a parallel, depth-synchronous incomplete alternating BFS on $\cG_{\cA}$ (Figure~\ref{fig:dsfm-parallel-frontier}).
The matching and admissible graph remain fixed during path generation and packing, until the selected batch is committed.
At depth $d$, column $j$ carries retained path labels $\cL_d(j)$, and the frontier is $\cF_d=\set{j\in\cC:\cL_d(j)\ne\emptyset}$.
A label records a free root, predecessor information, and score; $\pi$ also denotes its represented path, with $d$ edges outside $M$ ($\pi^+$) and $d$ edges in $M$ ($\pi^-$).
Initially $\cL_0(j)=\set{(j)}$ at free columns and is empty elsewhere; at positive depths, endpoints are matched columns.
Workers generate capped, ranked extensions in parallel across frontier columns.

For a partial or completed alternating path $\pi$, define
\begin{equation}
  \begin{aligned}
    \Delta_s(\pi)
      &=\sum_{e\in\pi^+}s(e)-\sum_{e\in\pi^-}s(e),\\
    \Delta_{\widetilde w}(\pi)
      &=\sum_{e\in\pi^+}\widetilde w(e)
        -\sum_{e\in\pi^-}\widetilde w(e).
  \end{aligned}
\end{equation}
Among labels arriving at the same vertex and depth, retention prefers smaller $\Delta_s$.
Each row and column retains at most one label per free root and at most \DSFMLabelsPerVertex{} roots over the entire invocation; retained root--vertex pairs are never reconsidered.
This favors root diversity because a compatible batch can augment each root only once.
Branch, label, and depth limits make the BFS incomplete: it can miss an augmenting path in $\cG_{\cA}$.

The depth limit $D$ counts added edges in a completed augmenting path.
Completed candidates accumulate in $\cT$ without further propagation.
Let $d_{\min}=\min_{\pi\in\cT}\card{\pi^+}$, initially $+\infty$; the first generated terminal depth fixes this value, which need not be the shortest augmenting-path depth in $\cG_{\cA}$.
Search continues only while
\begin{equation}
    d<\min\{D,d_{\min}+D_+\},\qquad D_+=\DSFMExtraDepth.\label{eq:dmin-update}
\end{equation}
The extra depths allow longer alternatives with lower slack or fewer conflicts; this rule may stop the search before $D$.

For a label $\pi$ ending at column $j$, {\bf $\Call{BranchEdges}{\pi,\cG_{\cA}}$} returns a capped set $B_\pi\subseteq\cA_j\setminus M$ of lowest-slack admissible edges.
Post-schedule searches also include edges minimizing the replacement-slack increment $s(i,j)-s(i,\mate_M(i))$, with the second term zero at free rows.
The immediate back-edge is excluded.
Larger branch or label limits preserve more alternatives at increased generation and storage cost, and potentially greater packing cost; tighter limits can shift work to completion.

{\bf $\Call{AdvanceDepth}{\cY_d,M}$} applies the invocation-wide retention rules in parallel over destination rows.
Retained extensions reaching free rows form $\cT_d$, with $d+1$ added and $d$ removed edges.
Others follow $(i,\mate_M(i))\in M$; parallel column-side retention forms $\cL_{d+1}$ and $\cF_{d+1}$ before the next depth begins.
It returns $(\cF_{d+1},\cT_d)$; labels and retention history remain invocation-local state.

{\bf $\Call{CandidateComponents}{\cT}$} retains at most \DSFMTerminalsPerRoot{} candidates per root, preferring smaller $\Delta_s$, subject to any invocation-wide candidate cap.
It groups them into maximal components connected by shared row or column vertices.
In Algorithm~\ref{alg:frontier}, $\pi\circ(i,j)$ extends $\pi$ to row $i$, and $\cY_{d,j}$ collects proposals from column $j$ before their union forms $\cY_d$.

\begin{algorithm}[!htb]
\caption{Score-guided frontier search}
\label{alg:frontier}
\small
\begin{algorithmic}[1]
\Require Current admissible graph $\cG_{\cA}$, matching $M$, and max search depth $D$
\Ensure Number of committed augmenting paths
\Procedure{FrontierSearch}{$\cG_{\cA},M$}
  \State $\cF_0\gets
    \set{j\in\cC:\mate_M(j)=\Nil}$
  \State $\cT\gets\emptyset$; $d_{\min}\gets\Infinity$

  \For{$d\gets0$ \textbf{to} $D-1$}
    \If{$\cF_d=\emptyset$ \textbf{or}
        $d\ge d_{\min}+D_+$}
      \textbf{break}
    \EndIf

    \ForAll{$j\in\cF_d$ \textbf{in parallel}}
      \State $\cY_{d,j}\gets\emptyset$
      \ForAll{labels $\pi\in\cL_d(j)$}
        \State $B_\pi\gets\Call{BranchEdges}{\pi,\cG_{\cA}}$
        \State $\cY_{d,j}\gets\cY_{d,j}\cup
        \set{\pi\circ(i,j):(i,j)\in B_\pi}$
        \EndFor
    \EndFor

    \State $\cY_d\gets\bigcup_{j\in\cF_d}\cY_{d,j}$ \Comment{collect proposals}
    \State $(\cF_{d+1},\cT_d)
      \gets\Call{AdvanceDepth}{\cY_d,M}$
    \If{$\cT_d\ne\emptyset$}
      \State $\cT\gets\cT\cup\cT_d$ \Comment{collect terminal paths}
      \State $d_{\min}\gets\min\{d_{\min},d+1\}$ \Comment{set on first $\cT\not=\emptyset$}
    \EndIf
    
  \EndFor

  \State $\cK\gets\Call{CandidateComponents}{\cT}$
  \If{$\cK=\emptyset$} \Return $0$
  \EndIf

  \ForAll{$K\in\cK$ \textbf{in parallel}}
    \If{$K$ is small} $S_K\gets\Call{CappedBnB}{K}$
    \Else ~$S_K\gets\Call{PriceProject}{K}$
    \EndIf
  \EndFor

  \State $S\gets\bigcup_{K\in\cK}S_K$ \Comment{commit disjoint paths in parallel}
  \State $ M\gets (M\setminus\bigcup_{\pi\in S}\pi^-) \cup \bigcup_{\pi\in S}\pi^+$
  \State \Return $\card{S}$
\EndProcedure
\end{algorithmic}
\end{algorithm}

\subsection{Conflict-aware path packing}
Candidate paths may share roots, internal vertices, or terminal rows and therefore cannot generally be committed together.
Packing selects a vertex-disjoint batch, preferring more paths and then smaller aggregate slack exchange: each selected path adds one to matching cardinality.
Let $\operatorname{res}(p)\subseteq\cC\cup\cR$ contain the vertices of path $p$; compatibility means disjoint resource sets.
Different conflict components are packed concurrently, and their selected batches are committed in parallel.

{\bf $\Call{CappedBnB}{K}$} is applied to components containing at most \DSFMPackingMaxRoots{} free roots, \DSFMPackingMaxPaths{} candidate paths, and \DSFMPackingMaxResources{} vertices.
Feasible batches use the cardinality-first ordering above.
Branch-and-bound explores at most \DSFMBnBNodeCap{} nodes, pruning subtrees that cannot improve the incumbent cardinality or, at equal cardinality, its aggregate slack change.
If the search completes within this limit, it returns the optimum under this ordering.
Otherwise, it returns the best feasible incumbent encountered and records that the node cap was exhausted.

Components exceeding any of these limits are handled by {\bf $\Call{PriceProject}{K}$}.
It uses a Lagrangian relaxation of the constraints requiring each row and column to occur in at most one selected path~\cite{Fisher1981,Held1974}.
Let $\cP_r(K)$ contain the candidate paths from free root $r$, and let $\mu_x^{(t)}\ge0$ denote the price associated with resource $x$ at iteration $t$.
Write $\cP(K)=\bigcup_r\cP_r(K)$ and define
\begin{equation}
  f_x=\card{\set{p\in\cP(K):x\in\operatorname{res}(p)}}.
\end{equation}
The unpriced cost of a candidate path is
\begin{equation}
  c_0(p)
  =
  \Delta_s(p)
  +
  \gamma
  \sum_{x\in\operatorname{res}(p)}\log f_x,
  \label{eq:path-base-cost}
\end{equation}
where $\gamma$ controls the static penalty assigned to edges and vertices shared by many candidates.
At iteration $t$, each root proposes a path minimizing
\begin{equation}
  c_t(p)
  =
  c_0(p)+
  \sum_{x\in\operatorname{res}(p)}\mu_x^{(t)},
  \qquad p\in\cP_r(K).
\end{equation}
If $\nu_x^{(t)}$ proposed paths use resource $x$, its price is updated by
\begin{equation}
  \mu_x^{(t+1)}
  =
  \pospart{
    \mu_x^{(t)}
    +
    \frac{\alpha}{\sqrt{t}}
    \bigl(\nu_x^{(t)}-1\bigr)
  }.
\end{equation}
Thus contention increases a resource's price, whereas the price of an unused resource decreases toward zero.
At prescribed iterations, DSFM projects the candidates under the current adjusted costs onto a feasible batch.
It returns the best feasible batch encountered, according to the batch
ordering above, during at most \DSFMCongestionIterations{} pricing iterations.
No approximation ratio is claimed for this packing heuristic.

\section{Certified completion}\label{sec:completion}
Bounded-search failure does not establish structural singularity.
Certified completion alternates parallel shortest-augmenting-path BFS~\cite{HopcroftKarp1973} on the current support with slack-ordered local exposure (Algorithm~\ref{alg:certified}).
BFS starts from every free column and completes the first depth reaching a free row; DSFM then augments along one or more shortest paths.
Only exhaustion without a free row triggers widening; then $M$ is maximum-cardinality on the admissible graph by Berge's theorem~\cite{Berge1957}.
Any full-graph augmenting path must leave the reached region through an unexposed edge at a reached column.
$\Call{Widen}{}$ exposes further edges there, retaining all earlier edges, then invokes the parallel $\Call{FrontierSearch}{}$.

The correction record $\rho$, initially zero, is the largest preceding finite $\sigma_{\mathrm{next}}$ from a failed complete search.
Only increases beyond $\rho$ contribute new corrections; $\rho$ is distinct from the exposure target $\lambda_{\mathrm{cert}}$, and search slacks remain fixed.

\begin{algorithm}[!htb]
\caption{Certified completion}
\label{alg:certified}
\small
\begin{algorithmic}[1]
\Require $\gG$, matching $M$, current admissible edge set $\cA$,
         $\omega=0$, and output copies initialized as in Algorithm~\ref{alg:dsfm}
\Ensure Returns $(H,\omega)$ with $\omega$ certified and output potentials feasible;
        either $M$ is perfect and $H=\Nil$, or $M$ is
        maximum-cardinality and $H$ is a Hall certificate
\Procedure{CertifiedCompletion}{$\gG,\cA,M,\omega$}
  \State $\rho\gets0$
  \State $\lambda_{\mathrm{cert}}\gets
    \max\set{s(e):e\in\cA}$
    \Comment{$0$ if $\cA=\emptyset$}
  \While{$M$ is not perfect}
    \State $(S,T,\ell,d)\gets
      \Call{AlternatingBFS}{\cG_{\cA},M}$
    \If{$d<\Infinity$}
      \State \Call{AugmentShortest}{$\cG_{\cA},M,\ell,d$}
    \Else
      \State $(\Delta\cA,\lambda_{\mathrm{cert}},
               \sigma_{\mathrm{next}})\gets\Call{Widen} {\gG,\cA,S,\lambda_{\mathrm{cert}}}$
      \If{$\Delta\cA\ne\emptyset$}
        \State $\alpha\gets\pospart{\sigma_{\mathrm{next}}-\rho}$
        \State $\omega\gets\omega+(\card{S}-\card{T})\alpha$
        \State $\rho\gets\max\{\rho,\sigma_{\mathrm{next}}\}$
        \If{$\alpha>0$}
          \State $u_i^{\mathrm{out}}\gets u_i^{\mathrm{out}}+\alpha$ for $i\in T$
          \State $v_j^{\mathrm{out}}\gets v_j^{\mathrm{out}}-\alpha$ for $j\in S$
        \EndIf
        \State \Call{FrontierSearch}{$\cG_{\cA},M$}
          \Comment{depth \DSFMLayerDepth}
      \Else
        \State \Return $((S,T),\omega)$
      \EndIf
    \EndIf
  \EndWhile
  \State \Return $(\Nil,\omega)$
\EndProcedure
\end{algorithmic}
\end{algorithm}

{\bf $\Call{AlternatingBFS}{\cG_{\cA},M}$} is an unpruned depth-synchronous implementation of BFS.
It returns reached columns $S$, rows $T$, BFS levels $\ell$, and minimum augmenting-path distance $d$ ($+\infty$ if none exists).
{\bf $\Call{AugmentShortest}{\cG_{\cA},M,\ell,d}$} uses these levels to pack and commit a guarded parallel batch of captured shortest paths.
If ineligible or rejected, a sequential blocking search guarantees at least one augmentation and may commit several.

If the alternating search terminates without reaching a free row, {\bf $\Call{Widen}{\gG,\cA,S,\lambda_{\mathrm{cert}}}$} finds
\begin{equation}
  \sigma_{\mathrm{next}} =
  \min\{s(i,j): j\in S,\, (i,j)\in\cE_j\setminus\cA\}.
  \label{eq:next-reached-slack}
\end{equation}
The minimum is evaluated by a reduction over the reached columns before the current expansion, using the fixed search slacks.
If no such edge exists, it returns $(\emptyset,\lambda_{\mathrm{cert}},\Nil)$ without changing the admissible sets or widening level.
If $\sigma_{\mathrm{next}}$ exists, the certified widening level advances to
\begin{equation}
  \lambda_{\mathrm{cert}}^+
  =
  \begin{cases}
    \sigma_{\mathrm{next}}, & \lambda_{\mathrm{cert}}=0,\\
    \max \{ \sigma_{\mathrm{next}},
      \DSFMWidenGrowth\lambda_{\mathrm{cert}}
    \},
    & \lambda_{\mathrm{cert}}>0.
  \end{cases}
\end{equation}
For every $j\in S$, define
\begin{equation}
    \begin{aligned}
        \cA_j^+ &= \cA_j\cup\cE_j^{\leq\lambda_{\mathrm{cert}}^+},\\
        \Delta\cA_j &= \cA_j^+\setminus\cA_j.
    \end{aligned}
\end{equation}
The operation replaces $\cA_j$ by $\cA_j^+$ in parallel over $j\in S$, leaves all other admissible sets unchanged, sets $\lambda_{\mathrm{cert}}\gets\lambda_{\mathrm{cert}}^+$, and returns $\lambda_{\mathrm{cert}}$, $\sigma_{\mathrm{next}}$, and
\begin{equation}
    \Delta\cA=\bigcup_{j\in S}\Delta\cA_j.
\end{equation}

If no unexposed edge remains at $S$, then $T=N_{\gG}(S)$ and $\card S-\card T=n-\card M>0$.
Thus $(S,T)$ is a Hall certificate and $M$ is maximum-cardinality in the full graph (Theorem~\ref{thm:analysis-completion}).

\section{Analysis of the parallel algorithm}\label{sec:analysis}
We analyze the parallel search, the objective information carried by fixed slack, and the structural guarantees of certified completion.
All statements use exact arithmetic and dual-feasible fixed slacks.
Empty neighborhoods are assumed to be handled before numerical normalization.
Throughout, let $m=\card{\cE}$.

\subsection{Parallel execution and bounded search}
For both the analysis and realized implementation, we use the phase-synchronous execution described in Sections~\ref{sec:overall}--\ref{sec:completion}.
Each depth resolves all proposals before publishing the next frontier.
Component solvers and the shortest-path fallback need not be internally parallel.
Within an obstruction correction, distinct row and column output entries can be updated independently; successive corrections are applied in search order.

\begin{restatable}[Safe and deterministic parallel transitions]{restatableproposition}{deterministic}\label{prop:analysis-parallel-transitions}
    Let $\mathcal B$ be a disjoint set of $M$-augmenting paths.
    Their simultaneous augmentation produces a matching $M'$ satisfying
    \begin{equation}
        \begin{aligned}
            \card{M'} &= \card{M} + \card{\mathcal{B}},\\
            s(M') &= s(M) + \sum_{p \in \mathcal{B}}\Delta_s (p),
        \end{aligned}
    \end{equation}
    i.e., the updates commute.

    DSFM computes precisely the same update independent of thread count and execution order provided the following conditions are met:
    \begin{enumerate}[label=(\roman*),noitemsep,topsep=4pt]
        \item each depth step resolves its complete logical proposal set by the prescribed total order;
        \item component solvers use deterministic search orders;
        \item and node, candidate, and workspace caps are allocated independently of thread-arrival order.
    \end{enumerate}
\end{restatable}

Let $q'$ be the maximum number of distinct roots retained at any vertex over one invocation, and let $f=n-\card{M}$ be the number of free columns at its start.
Set $q=\min\{q',f\}$.
Let $b_{\mathrm{br}}$ bound the number of edges returned by $\Call{BranchEdges}{}$;
the maximum depth $D$ bounds the number of added edges in a completed augmenting path.

\begin{restatable}[Bounds on retained labels and proposals]{restatableproposition}{frontierbound}\label{prop:analysis-frontier-bound}
One invocation retains labels for at most $2qn$ distinct root--vertex pairs, counting both row and column endpoints.
The total number of generated extension proposals satisfies
\begin{equation}
    \begin{aligned}
        N_{\mathrm{ext}}
        &=\sum_{d\ \mathrm{processed}}\card{\cY_d}\\
        &\le q\sum_{j\in\cC}
            \min\{b_{\mathrm{br}},\card{\cA_j}\}\\
        &\le q\min\{b_{\mathrm{br}}n,\card{\cA}\}.
    \end{aligned}
    \label{eq:analysis-extension-bound}
\end{equation}
If $t$ completed augmenting-path candidates are reconstructed, their total row--column incidence $I$ satisfies $I\le2Dt$.
\end{restatable}

The proposal bound is not a bound on all adjacency inspections.
Selecting replacement-slack candidates may require scanning an entire admissible column.
With at most one such scan per retained label, the inspection count is $\cO(q\card{\cA})$.
If per-vertex scratch arrays are initialized at every depth, this adds $\cO(Dn)$ work.
If components are solved sequentially but concurrently with one another (as in the realized implementation), their contribution to span is governed by the largest component solve.
Monotone exposure bounds newly admitted edges over the execution by $m$, but does not prevent repeated scans of already admissible edges.

\subsection{Fixed-slack quality and support}\label{sec:fixed-slack-quality}
Assume a perfect matching exists, then write
\begin{equation}
    \begin{aligned}
        \mathcal D&=\sum_i u_i+\sum_jv_j,\\
        \delta_{\mathrm{fix}}
        &=\mathcal D-w(M^\star)
        =\min_{N\ \mathrm{perfect}}s(N).
    \end{aligned}
\end{equation}
The fixed-slack identity gives, for every perfect $M$,
\begin{equation}
 w(M^\star)-w(M)=s(M)-\delta_{\mathrm{fix}}.
 \label{eq:analysis-exact-gap}
\end{equation}
If $M\subseteq\cA$, then
\begin{equation}
 w(M^\star)-w(M)
 \le s(M)
 \le\sum_{j\in\cC}\max_{e\in\cA_j}s(e).
\end{equation}
If every exposed edge has slack at most $\lambda$, the geometric-mean quality bound is at least $e^{-\lambda}$.

For analysis, define the uniform fixed-slack layer $\cE^{\le\eta}=\bigcup_j\cE_j^{\le\eta}$.
For any perfect reference matching $M^p$, set
\begin{equation}
    Z_{\mathrm{off}}(M^p) = \sum_{e \in \cE \setminus M^p}e^{-s(e)}.
\end{equation}

\begin{restatable}[Edge concentration and layer size]{restatableproposition}{edgeconcentration}\label{prop:analysis-support}
    For every $\eta>0$, every maximum-weight perfect matching $M^\star$ and every perfect reference matching $M^p$ satisfy
    \begin{equation}
        \begin{aligned}
            \card{M^\star \setminus \cE^{\le \eta}}
            &\le \frac{\delta_{\mathrm{fix}}}{\eta} \le \frac{s(M^p)}{\eta},\\
            \card{\cE^{\le \eta}}
            &\le n+e^\eta Z_{\mathrm{off}}(M^p).
        \end{aligned}
    \end{equation}
    These statements are fully independent of Sinkhorn (Algorithm~\ref{alg:sinkhorn}) convergence.
\end{restatable}

The two inequality quantify two properties important to DSFM: how much optimal-matching support lies in a layer, and how many competing graph edges occupy it.
Neither guarantees a small layer without a numerical concentration assumption.

\subsection{Certified completion and unavoidable slack}
Certified completion uses exhaustive alternating searches primarily to guide edge exposure, but also exposes the fixed slack required by any perfect matching.
The resulting Hall obstructions therefore support both structural completeness and a tighter a posteriori objective certificate.

\begin{restatable}[Hall obstruction, slack cost, and completion]{restatabletheorem}{hallobstruction}\label{thm:analysis-completion}
    Suppose $M$ is non-perfect and complete alternating BFS in $\cG_{\cA}$ reaches no free row.
    Let $S,T$ be its reached columns and rows, then
    \begin{equation}
        T=N_{\cG_{\cA}}(S),\qquad h:=\card{S}-\card{T}=n-\card{M}>0,
    \end{equation}
    and $M$ is maximum-cardinality in the admissible graph.
    Every perfect matching in the full graph contains at least $h$ unexposed edges from $S$ to $\cR\setminus T$.
    Whenever $\sigma_{\mathrm{next}}$ exists (equation~\ref{eq:next-reached-slack}),
    \begin{equation}
     \delta_{\mathrm{fix}}\ge h\sigma_{\mathrm{next}},
     \label{eq:analysis-hall-price}
    \end{equation}
    given a perfect matching exists in the full graph.

    Certified completion terminates with either a perfect matching or a maximum-cardinality matching and Hall certificate satisfying $\card S-\card T=n-\card M$.
    If it begins with matching $M_{\mathrm{in}}$ and admissible set $\cA_{\mathrm{in}}$, and performs $J$ nonempty widenings, then
    \begin{equation}
        \begin{aligned}
            N_{\mathrm{BFS}}&\le n-\card{M_{\mathrm{in}}}+J+1,\\
            J&\le m-\card{\cA_{\mathrm{in}}}.
        \end{aligned}
    \end{equation}
\end{restatable}

Equation~\eqref{eq:analysis-hall-price} can be strengthened by accumulating distinct certified slack intervals rather than selecting one obstruction.
Index the failed complete searches with finite $\sigma_{\mathrm{next}}$ by $t=1,\ldots,J$, and write $\sigma_t=\sigma_{\mathrm{next},t}$ and $h_t=\card{S_t}-\card{T_t}$.
With $\rho_0=\omega_0=0$, Algorithm~\ref{alg:certified} computes
\begin{equation}
    \begin{aligned}
        \rho_t&=\max\{\rho_{t-1},\sigma_t\},\\
        \alpha_t&=\rho_t-\rho_{t-1},\\
        \omega_t&=\omega_{t-1}+h_t\alpha_t,
        \qquad \omega=\omega_J.
    \end{aligned}
    \label{eq:accumulated-correction}
\end{equation}
Only newly certified slack intervals are charged, avoiding repeated charges for overlapping obstructions.

\begin{restatable}[Accumulated obstruction corrections]{restatableproposition}{obstructioncorrection}
\label{prop:accumulated-corrections}
    Before optional final tightening, the accumulated output potentials remain dual feasible and satisfy
    \begin{equation}
        \sum_i u_i^{\mathrm{out}}+\sum_j v_j^{\mathrm{out}}
        =\mathcal D-\omega.
        \label{eq:accumulated-output-objective}
    \end{equation}
    If a perfect matching exists, then
    \begin{equation}
        \max\bigl(\{0\}\cup\{h_t\sigma_t:1\le t\le J\}\bigr)
        \le\omega\le\delta_{\mathrm{fix}}.
        \label{eq:accumulated-dominance}
    \end{equation}
    Thus accumulation is never weaker than the largest individual obstruction bound.
\end{restatable}

For a returned perfect matching $M_f$, $0\le\omega\le\delta_{\mathrm{fix}}\le s(M_f)$.
he certificate of Section~\ref{sec:dual-cert} therefore improves the matched-slack geometric-mean bound by $\exp(\omega/n)$; equality $s(M_f)=\omega$ certifies optimality even with positive matched slack.

\subsection{Worst-case work model}\label{sec:concentration-work}
Let $m=\card{\cE}$, and let $a$ and $b$ be the admissible edge counts at the end of scheduled exposure and at termination.
The post-schedule searches do not expose edges, so $a$ is also the count at completion entry.
Let $h=n-\card{M_{\mathrm{in}}}$ be the entry matching's shortfall and $J$ the number of nonempty certified widenings.
Completion contributes zero work when it is not entered.

We count aggregate RAM work with unit-cost weight operations and fixed reference budgets.
We assume ordered-prefix exposure, linear-work seed arbitration, comparison-based frontier grouping, and incidence-based conflict construction.
With $d_j=\card{\cE_j}$, full-graph preprocessing costs
\begin{equation}
 W_{\mathrm{pre}}
 =\cO\!\bigl(m+n+\sum_j d_j\log(d_j+1)\bigr).
 \label{eq:work-preprocessing}
\end{equation}

For pricing/projection, let $t$ count the completed candidate paths being processed and let $I$ be their total row--column incidence.
Pricing and each proposal--winner projection round scan these candidate and incidence arrays in $\cO(I+t)$ work.
With a fixed number of projection rounds, retaining the best batch, including sorting selected candidate identifiers, gives $\cO(I+t\log(t+1))$ work per projection checkpoint.
The bounded number of pricing iterations and checkpoints preserves this bound for the pricing/projection stage.
Certified BFS and shortest-path augmentation, including any captured-path attempt, use $\cO(n+\card{\cA})$ work per iteration under this model.

The schedule and the post-schedule phase (at most \DSFMPostRounds{} searches) each cost $\cO(a+n\log(n+1))$.
Shallow searches generate $\cO(n)$ extension proposals and reconstruct $\cO(n)$ bounded-length augmenting-path candidates; deep searches additionally scan admissible columns for replacement-slack selection.
Completion makes at most $h+J+1$ BFS calls and invokes one shallow search per nonempty widening.
Thus
\begin{equation}
    \begin{aligned}
        W_{\mathrm{sched}}+W_{\mathrm{post}}
        &=\cO(a+n\log(n+1)),\\
        W_{\mathrm{cert}}
        &=\cO\bigl((h+J+1)(n+b)\\
        &\qquad +Jn\log(n+1)\bigr).
    \end{aligned}
\label{eq:work-phases}
\end{equation}
Since $J\le b-a$, $h\le n$, and $a\le b\le m$, the conservative total bound is
\begin{equation}
 W=\cO\!\bigl(W_{\mathrm{pre}}+(n+m)[m+n\log(n+1)]\bigr).
 \label{eq:work-worst}
\end{equation}

\section{Empirical Results}\label{sec:experiments}

We select from the SuiteSparse Matrix Collection~\cite{Davis2011} all non-diagonal square real matrices with dimension greater than $10{,}000$, where ``Rutherford--Boeing Type'' is marked as \texttt{real}.
We exclude only \texttt{MOLIERE\_2016}, whose number of nonzeros exceeds the range supported by 32-bit sparse indices.
The resulting set contains 617 matrices: $298$ unsymmetric and $319$ symmetric.
Of these, 598 admit a perfect matching and have a validated MC64 reference assignment; the remaining 19 are structurally singular and are reported only in the termination-stage summary.

We compare DSFM with MC64 and HWPM.
For parallel methods, each matrix is run at several thread or rank counts, with three repetitions at each configuration.
For each method, we use the configuration median having the smallest median preprocessing time, and report matching quality from that same configuration.
Preprocessing time includes input normalization, matching, certificate construction, and returned scaling for DSFM;
input conversion, workspace preparation, matching/scaling, and result processing for MC64;
and distributed matrix construction, matching, gathering, and result processing for HWPM.
All three start from resident input and exclude file loading, output serialization, and additional diagnostics.
Experiments were performed on a CPU-only compute node of NERSC's Perlmutter system, with two AMD Epyc 7763 processors, each with 64 physical cores, and 256~GB of memory per socket.

\subsection{Matching quality, matrix scaling, and certification}\label{sec:matching-certificate}
\begin{figure}[!htb]
  \centering
  \includegraphics[width=\linewidth]{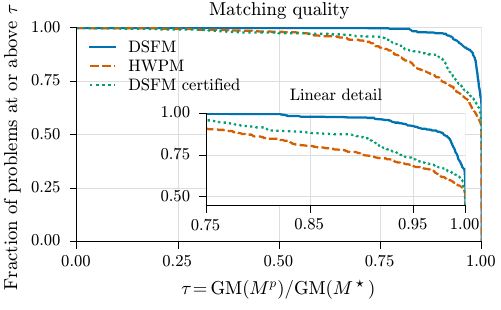}
  \caption{Fraction of the 598 perfect-reference problems attaining
  MC64-relative geometric-mean quality at least $\tau$.
  The ``certified'' curve is DSFM's returned quality lower bound.
  The inset enlarges the high-quality region.}
  \label{fig:empirical-quality}
\end{figure}

Figure~\ref{fig:empirical-quality} measures aggregate product quality, but the same objective loss can be distributed differently among the selected entries.
Across nearly the entire nontrivial threshold range, DSFM attains higher MC64-relative matching quality than HWPM, with the largest separation in the high-quality regime.
To examine this distribution independently of each method's returned scaling, we apply MC64's scaling to both matchings.
In exact arithmetic, MC64's matched entries have unit magnitude and all entries are bounded by one, providing a common unit benchmark in every row and column.
The minimum and lower tail show whether the objective loss relative to MC64 is spread across many entries or concentrated in a few particularly weak selections.

Under this common scaling, DSFM's minimum matched magnitude is better, equal ($10^{-8}$ tolerance), or worse than HWPM's on 288, 262, and 48 problems, respectively.
Nevertheless, among the 297 problems where DSFM has higher geometric-mean matching quality, its minimum is worse on 15.
Higher product quality therefore does not guarantee improvement in
the weakest selected entry.

\begin{figure}[!htb]
  \centering
  \includegraphics[width=\linewidth]{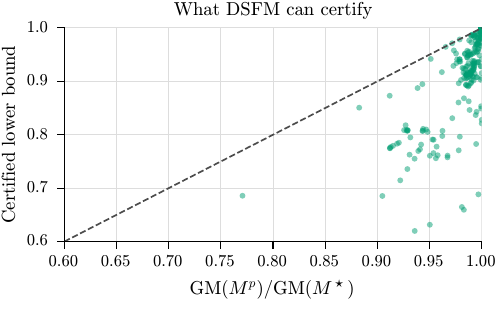}
  \caption{Returned DSFM lower bound versus observed MC64-relative quality, shown in detail on $[0.60,1]^2$.}
  \label{fig:empirical-certificate}
\end{figure}

Figure~\ref{fig:empirical-certificate} shows how conservative DSFM's a posteriori quality guarantee is relative to the observed matching quality.
On the 140 problems where the Hall correction $\omega$ is positive, it closes a median 17.96\% of the gap between the fixed-dual upper bound and the MC64 reference objective, measured in log-weight coordinates.

\subsection{Admissible support and completion}

\begin{table}[!htb]
  \centering
  \includegraphics[width=\linewidth]{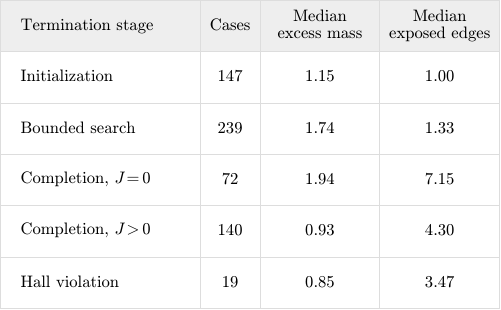}
    \caption{Termination stage across 617 matrices.
    Excess mass is
    $1/n\sum_{e\in\cE}\exp(-s(e))-1$.
    Exposed edges is the column-average admissible edges after bounded search.}
  \label{tab:empirical-exit-support}
\end{table}

Table~\ref{tab:empirical-exit-support} supports the concentration premise behind DSFM.
Of the 598 nonsingular problems, 386 finish during initialization or during the bounded searches; these two groups expose a median of only $1.00$ and $1.33$ edges per column, respectively.
Thus, on most problems, a perfect matching is recovered from a small numerically selected subset of the matrix graph rather than from the full support.
The remaining cases illustrate why DSFM does not purely rely on concentration: 212 problems enter certified completion, of which 140 require further edge exposure.

\begin{figure}[!htb]
  \centering
  \includegraphics[width=\linewidth]{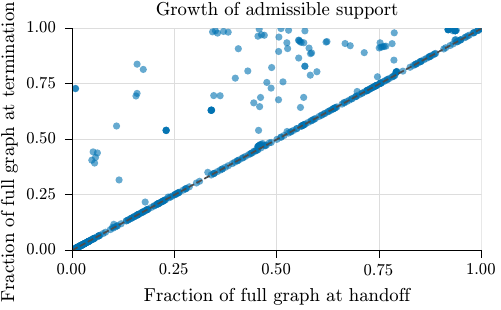}
  \caption{Admissible support fractions before completion and at termination.
  Early exits use their exit support for both coordinates.
  Diagonal points include completion runs requiring no additional exposure.}
  \label{fig:empirical-support}
\end{figure}

Figure~\ref{fig:empirical-support} shows how this fallback enlarges the active graph when the bounded search is insufficient.
Many runs lie on the diagonal, including 72 nonsingular completion runs whose handoff support already admits a perfect matching.
Other problems require progressively larger supports, in some cases reaching the full graph.
Thus the active-subgraph policy exploits concentration when it is present while certified completion supplies the additional structure needed when it is not.

\subsection{Preprocessing time, strong-scaling, and quality tradeoffs}

We consider the 37 perfect-reference matrices whose median MC64
preprocessing time is at least one second, focusing the runtime
comparison on the more expensive instances.
The focused set spans all DSFM exit modes: 12 problems finish during
initialization, 9 during bounded search, 9 enter certified completion without further exposure, and 7 require certified widening.
Thus 16 of the 37 problems exercise certified completion.
On the remaining 561 matrices, the largest preprocessing times are 0.96, 2.50, and 7.63 seconds for MC64, DSFM, and HWPM, respectively.

\begin{figure}[!htb]
  \centering
  \includegraphics[width=\linewidth]{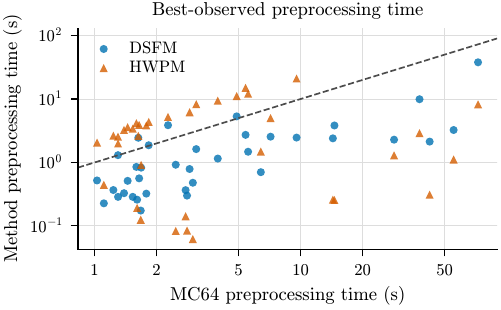}
  \caption{Best-observed preprocessing time on the 37 perfect-reference matrices with MC64 median time at least one second.
  Each method selects its own fastest.}
  \label{fig:empirical-runtime}
\end{figure}

Across these 37 problems, DSFM and HWPM achieve geometric-mean speedups over MC64 of $3.53\times$ and $2.29\times$, respectively.
Pairing the runtimes in Figure~\ref{fig:empirical-runtime} with matching quality, DSFM is faster with no worse quality than HWPM on 23 of the 37 problems, whereas HWPM has this advantage on six.
Among the ten problems where HWPM is at least twice as fast, five have geometric-mean quality more than one percentage point below DSFM's.

\begin{figure}[!htb]
  \centering
  \includegraphics[width=\linewidth]{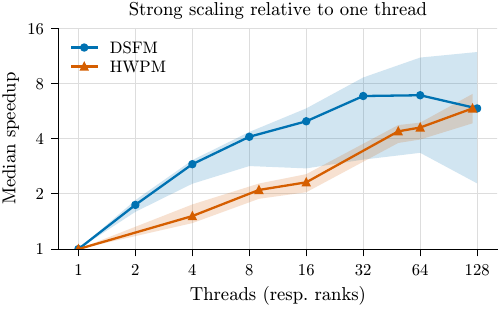}
    \caption{Strong scaling on the same 37 reference matrices.
    Speedup is relative to one thread for DSFM and one MPI rank for HWPM.
    Curves show the median across matrices and shading shows the
    IQR.}
  \label{fig:strong-scaling}
\end{figure}

Figure~\ref{fig:strong-scaling} shows that DSFM scales more strongly than HWPM over most of the tested worker range.
Although the median DSFM speedup begins to level off at high thread counts, the broad interquartile range shows substantial matrix-to-matrix variation, with the upper quartile continuing to obtain large speedups.
We attribute the eventual saturation primarily to irregular cross-socket accesses to row data: first-touch allocation improves locality, but sparse column traversals cannot keep shared row state local to a single NUMA domain.

\section{Discussion \& Conclusion}\label{sec:conclusion}

DSFM combines fixed-slack active-subgraph search with structural completion, objective certification, and matching-based scaling.
Accumulated Hall obstructions strengthen both the quality guarantee and the returned scaling without changing the search slacks.
On 598 perfect-reference problems, DSFM attains higher matching quality than HWPM over most quality thresholds.
On the 37 problems requiring at least one second of MC64 preprocessing, DSFM achieves a $3.53\times$ geometric-mean speedup using each matrix's best observed configuration.
These results support DSFM as a quality--performance alternative rather than a uniformly faster replacement.
End-to-end evaluation in 
a sparse direct solver remains a next step for assessing how these preprocessing gains affect factorization robustness and total solution time.

\section*{Acknowledgments.}
This work was supported in part by the the U.S. Department of Energy, Office of Science,
Office of Advanced Scientific Computing Research, Mathematical Multifaceted Integrated
Capability Centers (MMICCs) through the Sparsitute Center, under contract number DE-AC02-05CH11231.

\bibliographystyle{siamplain}
\bibliography{references}

@article{Duff2001,
  title = {On Algorithms For Permuting Large Entries to the Diagonal of a Sparse Matrix},
  volume = {22},
  ISSN = {1095-7162},
  url = {http://dx.doi.org/10.1137/S0895479899358443},
  DOI = {10.1137/s0895479899358443},
  number = {4},
  journal = {SIAM Journal on Matrix Analysis and Applications},
  publisher = {Society for Industrial & Applied Mathematics (SIAM)},
  author = {Duff,  I. S. and Koster,  J.},
  year = {2001},
  month = Jan,
  pages = {973–996}
}

@article{Duff1999,
  title = {The Design and Use of Algorithms for Permuting Large Entries to the Diagonal of Sparse Matrices},
  volume = {20},
  ISSN = {1095-7162},
  url = {http://dx.doi.org/10.1137/S0895479897317661},
  DOI = {10.1137/s0895479897317661},
  number = {4},
  journal = {SIAM Journal on Matrix Analysis and Applications},
  publisher = {Society for Industrial & Applied Mathematics (SIAM)},
  author = {Duff,  Iain S. and Koster,  Jacko},
  year = {1999},
  month = Jan,
  pages = {889–901}
}

@article{Kuhn1955,
  title = {The Hungarian method for the assignment problem},
  volume = {2},
  ISSN = {1931-9193},
  url = {http://dx.doi.org/10.1002/nav.3800020109},
  DOI = {10.1002/nav.3800020109},
  number = {1-2},
  journal = {Naval Research Logistics Quarterly},
  publisher = {Wiley},
  author = {Kuhn,  H. W.},
  year = {1955},
  month = Mar,
  pages = {83–97}
}

@article{Sinkhorn1967,
  title = {Concerning nonnegative matrices and doubly stochastic matrices},
  volume = {21},
  ISSN = {0030-8730},
  url = {http://dx.doi.org/10.2140/pjm.1967.21.343},
  DOI = {10.2140/pjm.1967.21.343},
  number = {2},
  journal = {Pacific Journal of Mathematics},
  publisher = {Mathematical Sciences Publishers},
  author = {Sinkhorn,  Richard and Knopp,  Paul},
  year = {1967},
  month = May,
  pages = {343–348}
}

@article{Knight2008,
  title = {The Sinkhorn–Knopp Algorithm: Convergence and Applications},
  volume = {30},
  ISSN = {1095-7162},
  url = {http://dx.doi.org/10.1137/060659624},
  DOI = {10.1137/060659624},
  number = {1},
  journal = {SIAM Journal on Matrix Analysis and Applications},
  publisher = {Society for Industrial & Applied Mathematics (SIAM)},
  author = {Knight,  Philip A.},
  year = {2008},
  month = Jan,
  pages = {261–275}
}

@article{Hall1935,
  title = {On Representatives of Subsets},
  volume = {s1-10},
  ISSN = {0024-6107},
  url = {http://dx.doi.org/10.1112/jlms/s1-10.37.26},
  DOI = {10.1112/jlms/s1-10.37.26},
  number = {1},
  journal = {Journal of the London Mathematical Society},
  publisher = {Wiley},
  author = {Hall,  P.},
  year = {1935},
  month = Jan,
  pages = {26–30}
}

@article{Duff2005,
  title = {Strategies for Scaling and Pivoting for Sparse Symmetric Indefinite Problems},
  volume = {27},
  ISSN = {1095-7162},
  url = {http://dx.doi.org/10.1137/04061043X},
  DOI = {10.1137/04061043x},
  number = {2},
  journal = {SIAM Journal on Matrix Analysis and Applications},
  publisher = {Society for Industrial & Applied Mathematics (SIAM)},
  author = {Duff,  Iain S. and Pralet,  Stéphane},
  year = {2005},
  month = Jan,
  pages = {313–340}
}

@article{Held1974,
  title = {Validation of subgradient optimization},
  volume = {6},
  ISSN = {1436-4646},
  url = {http://dx.doi.org/10.1007/BF01580223},
  DOI = {10.1007/bf01580223},
  number = {1},
  journal = {Mathematical Programming},
  publisher = {Springer Science and Business Media LLC},
  author = {Held,  Michael and Wolfe,  Philip and Crowder,  Harlan P.},
  year = {1974},
  month = Dec,
  pages = {62–88}
}

@article{Fisher1981,
  title = {The Lagrangian Relaxation Method for Solving Integer Programming Problems},
  volume = {27},
  ISSN = {1526-5501},
  url = {http://dx.doi.org/10.1287/mnsc.27.1.1},
  DOI = {10.1287/mnsc.27.1.1},
  number = {1},
  journal = {Management Science},
  publisher = {Institute for Operations Research and the Management Sciences (INFORMS)},
  author = {Fisher,  Marshall L.},
  year = {1981},
  month = Jan,
  pages = {1–18}
}

@article{Davis2011,
  title = {The university of Florida sparse matrix collection},
  volume = {38},
  ISSN = {1557-7295},
  url = {http://dx.doi.org/10.1145/2049662.2049663},
  DOI = {10.1145/2049662.2049663},
  number = {1},
  journal = {ACM Transactions on Mathematical Software},
  publisher = {Association for Computing Machinery (ACM)},
  author = {Davis,  Timothy A. and Hu,  Yifan},
  year = {2011},
  month = Nov,
  pages = {1–25}
}

@article{Azad2020,
  title = {A Distributed-Memory Algorithm for Computing a Heavy-Weight Perfect Matching on Bipartite Graphs},
  volume = {42},
  ISSN = {1095-7197},
  url = {http://dx.doi.org/10.1137/18M1189348},
  DOI = {10.1137/18m1189348},
  number = {4},
  journal = {SIAM Journal on Scientific Computing},
  publisher = {Society for Industrial & Applied Mathematics (SIAM)},
  author = {Azad,  Ariful and Bulu\c{c},  Aydin and Li,  Xiaoye S. and Wang,  Xinliang and Langguth,  Johannes},
  year = {2020},
  month = Jan,
  pages = {C143–C168}
}

@book{CLRS2022,
  author    = {Cormen, Thomas H. and Leiserson, Charles E.
               and Rivest, Ronald L. and Stein, Clifford},
  title     = {Introduction to Algorithms},
  edition   = {4},
  publisher = {The MIT Press},
  address   = {Cambridge, Massachusetts},
  year      = {2022},
  isbn      = {9780262046305}
}

@misc{Sharify2013,
  doi = {10.48550/ARXIV.1104.3830},
  url = {https://arxiv.org/abs/1104.3830},
  author = {Sharify,  Meisam and Gaubert,  Stéphane and Grigori,  Laura},
  title = {Solution of the optimal assignment problem by diagonal scaling algorithms},
  publisher = {arXiv},
  year = {2011},
  copyright = {arXiv.org perpetual,  non-exclusive license}
}

@article{Azad2017,
  title = {Computing Maximum Cardinality Matchings in Parallel on Bipartite Graphs via Tree-Grafting},
  volume = {28},
  ISSN = {1045-9219},
  url = {http://dx.doi.org/10.1109/TPDS.2016.2546258},
  DOI = {10.1109/tpds.2016.2546258},
  number = {1},
  journal = {IEEE Transactions on Parallel and Distributed Systems},
  publisher = {Institute of Electrical and Electronics Engineers (IEEE)},
  author = {Azad,  Ariful and Buluc,  Aydn and Pothen,  Alex},
  year = {2017},
  month = Jan,
  pages = {44–59}
}

@article{Hook2019,
  title = {Max-Balanced Hungarian Scalings},
  volume = {40},
  ISSN = {1095-7162},
  url = {http://dx.doi.org/10.1137/15M1024871},
  DOI = {10.1137/15m1024871},
  number = {1},
  journal = {SIAM Journal on Matrix Analysis and Applications},
  publisher = {Society for Industrial & Applied Mathematics (SIAM)},
  author = {Hook,  James and Pestana,  Jennifer and Tisseur,  Fran\c{c}oise and Hogg,  Jonathan},
  year = {2019},
  month = Jan,
  pages = {320–346}
}

@article{HopcroftKarp1973,
  title = {An $n^{5/2} $ Algorithm for Maximum Matchings in Bipartite Graphs},
  volume = {2},
  ISSN = {1095-7111},
  url = {http://dx.doi.org/10.1137/0202019},
  DOI = {10.1137/0202019},
  number = {4},
  journal = {SIAM Journal on Computing},
  publisher = {Society for Industrial & Applied Mathematics (SIAM)},
  author = {Hopcroft,  John E. and Karp,  Richard M.},
  year = {1973},
  month = Dec,
  pages = {225–231}
}

@article{Berge1957,
  title = {TWO THEOREMS IN GRAPH THEORY},
  volume = {43},
  ISSN = {1091-6490},
  url = {http://dx.doi.org/10.1073/pnas.43.9.842},
  DOI = {10.1073/pnas.43.9.842},
  number = {9},
  journal = {Proceedings of the National Academy of Sciences},
  publisher = {National Academy of Sciences},
  author = {Berge,  Claude},
  year = {1957},
  month = Sep,
  pages = {842–844}
}

\clearpage

\appendix

\newcommand{\suppsection}[1]{\section*{\large #1}\vspace{-3pt}}
\setlength{\columnsep}{10pt}
\setlength{\parindent}{0pt}
\setlength{\parskip}{4pt}
\setlength{\multicolsep}{7pt}
\setlength{\abovecaptionskip}{5pt}
\setlength{\belowcaptionskip}{3pt}
\renewcommand{\thefigure}{S.\arabic{figure}}
\renewcommand{\thetable}{S.\arabic{table}}

\suppsection{A. Theoretical analysis with proofs}

\deterministic*

\begin{proof}
    Each internal path vertex loses one matched edge and gains one; each free endpoint gains one edge.
    Disjoint paths modify disjoint vertices, so their matching updates commute and their cardinality and slack changes are additive.
    Under the stated execution conditions, each phase produces a uniquely determined proposal set, retained labels and frontier, and feasible batch from its input state.
    Induction over phases gives determinism independent of execution schedule.
\end{proof}

\frontierbound*

\begin{proof}
Each vertex retains at most one label per root and labels from at most $q$ distinct roots over the invocation.
Summing over the $2n$ vertices gives the retained-pair bound.
Each retained label at column $j$ is expanded at most once, generating at most $\min\{b_{\mathrm{br}},\card{\cA_j}\}$ proposals.
There are at most $q$ such retained labels per column, giving the proposal bound.
A completed augmenting path has at most $D$ added and $D-1$ removed edges, hence at most $2D$ vertices.
Summing over the reconstructed candidates gives $I\le2Dt$.
\end{proof}

\edgeconcentration*

\begin{proof}
    The slacks of $M^\star$ sum to $\delta_\mathrm{fix} \le s(M^p)$ by definition, so at most $\delta_\mathrm{fix}/\eta$ of its edges can have slack greater than $\eta$.
    Each edge outside $M^p$ with slack at most $\eta$ contributes at least $e^{-\eta}$ to $Z_\mathrm{off}(M^p)$.
\end{proof}

\hallobstruction*

\begin{proof}
    Exhaustive alternating reachability gives the neighbor equality.
    Since no reached row is free, matched edges biject $T$ with the nonroot columns of $S$, giving the deficiency identity.
    This deficiency bounds any active-graph matching by $n-h=\card M$.

    At most $\card T$ columns of $S$ can be assigned to rows in $T$ by any perfect matching.
    At least $h$ must use rows outside $T$.
    These edges are unexposed, since otherwise BFS would have reached their rows, and each has slack at least $\sigma_{\mathrm{next}}$.
    Nonnegativity of the remaining slacks gives~\eqref{eq:analysis-hall-price}.

    Every successful BFS causes at least one augmentation.
    Every nonterminal failed BFS exposes at least one previously unexposed edge.
    There are at most $n-\card{M_{\mathrm{in}}}$ successful BFS calls, at most $J$ nonempty widenings, and at most one terminal failure.
    If no unexposed edge remains at the reached columns, their active neighbor set equals their full-graph neighbor set.
    The same deficiency then bounds all full-graph matchings by $\card M$, proving maximum cardinality and structural singularity.
    If a perfect matching exists, this Hall-certificate outcome is impossible.
\end{proof}

\obstructioncorrection*

\begin{proof}
    The total correction amplitude before search $t$ is $\sum_{r<t}\alpha_r=\rho_{t-1}$.
    Each earlier correction reduces any edge slack by at most its amplitude, so the current output slack is at least $s(e)-\rho_{t-1}$.
    The next correction reduces slack only on edges from $S_t$ to $\cR\setminus T_t$.
    Each has fixed slack at least $\sigma_t$ by Theorem~\ref{thm:analysis-completion}.
    When $\alpha_t>0$, its current output slack is therefore at least $\sigma_t-\rho_{t-1}=\alpha_t$, so the correction preserves feasibility.
    Other edge slacks do not decrease.

    The change in the output potential sum is $(\card{T_t}-\card{S_t})\alpha_t=-h_t\alpha_t$.
    Summing proves~\eqref{eq:accumulated-output-objective}; feasibility then gives $\omega\le\delta_{\mathrm{fix}}$.
    All searches start from every free column, so $h_t=n-\card{M_t}$ is nonincreasing, where $M_t$ is the matching at search $t$.
    Consequently, $\omega_t=\sum_{r\le t}h_r\alpha_r\ge h_t\rho_t\ge h_t\sigma_t$.
    Since $\omega_t$ is nondecreasing, the final $\omega$ dominates every individual bound.
\end{proof}

\end{document}